\documentclass[prD,superscriptaddress,twocolumn,showpacs,nofootinbib,amsmath,amssymb]{revtex4-1}
\usepackage{graphicx,bm,adjustbox}
\usepackage{color}
\usepackage{float, subcaption,caption,placeins}

\begin{document}
\pagenumbering{arabic}

\title{Impact of axisymmetric mass models for dwarf spheroidal galaxies
on indirect dark matter searches}

\author{Niki Klop}
\email{l.b.klop@uva.nl}
\affiliation{GRAPPA Institute, University of Amsterdam, 1098 XH
Amsterdam, The Netherlands}
\author{Fabio Zandanel}
\affiliation{GRAPPA Institute, University of Amsterdam, 1098 XH
Amsterdam, The Netherlands}
\author{Kohei Hayashi}
\affiliation{Kavli Institute for the Physics and Mathematics of the Universe (Kavli IPMU), The University of Tokyo, Chiba 277-8583, Japan}
\affiliation{Kavli Institute for Astronomy and Astrophysics (KIAA), Peking University, Beijing 100871, China}
\author{Shin'ichiro Ando}
\affiliation{GRAPPA Institute, University of Amsterdam, 1098 XH
Amsterdam, The Netherlands}
\date{\today}
\begin{abstract}

Dwarf spheroidals are low-luminosity satellite galaxies of the Milky Way
highly dominated by dark matter (DM). Therefore, they are prime targets to
search for signals from dark matter annihilation using gamma-ray
observations. While the typical assumption is that the dark matter density
profile of these satellite galaxies can be described by a spherical symmetric 
Navarro-Frenk-White (NFW) profile, recent observational data of stellar kinematics
suggest that the DM halos around these galaxies are better described by axisymmetric profiles. Motivated by such evidence, we analyse about seven years 
of \texttt{PASS8} \emph{Fermi} data for seven classical dwarf galaxies, including Draco, 
adopting both the widely used NFW profile and observationally-motivated axisymmetric 
density profiles. For four of the selected dwarfs (Sextans, Carina, Sculptor and Fornax) 
axisymmetric mass models suggest a cored density profile rather than the commonly adopted
cusped profile. We found that upper limits on the annihilation cross section for some of these 
dwarfs are significantly higher than the ones achieved using an NFW profile. Therefore, 
upper limits in the literature obtained using spherical symmetric cusped profiles, 
such as the NFW, might be overestimated. Our results show that it is extremely important 
to use observationally motivated density profiles going beyond the usually adopted NFW 
in order to obtain accurate constraints on the dark matter annihilation cross section.
 
\end{abstract}
\maketitle

\section{Introduction}
\label{sec:intro}

Most of the matter in the Universe consists of an unknown component
that is commonly considered to be made of non-baryonic cold dark matter
\cite{2011ApJS..192...18K,Ade:2015xua}.
Finding the particle nature of dark matter (DM) is one of the most pressing
goals in modern physics.
While many particle physics models have been proposed to solve this
puzzle, the most favored and extensively studied candidates fall into
the category of weakly interacting massive particles (WIMPs)
\cite{2005PhR...405..279B}.
These are characterised by a relic density matching the observed DM
density, and naturally arise in many theories beyond the standard model
of particle physics such as supersymmetry or universal extra-dimension
models.
The self-annihilation of WIMPs can result in the production of standard
model particles.
The goal of so-called indirect DM searches is to look for these
particles in regions of the Universe where we know DM is abundant
\cite{Gaskins:2016cha}.

High-energy gamma rays are one example of those particles expected as a result of WIMP annihilation.
The search for these gamma rays is a very active field of research
fueled in the last decade by many gamma-ray observations of Milky Way (MW) satellite galaxies
\cite{2009ApJ...697.1299A,2010ApJ...720.1174A,2011JCAP...06..035A,2011PhRvL.107x1302A,2012PhRvD..85f2001A,2014JCAP...02..008A,Abdo:2010ex,Ackermann:2013yva,2014PhRvD..90k2012A,Geringer-Sameth:2014qqa,Ackermann:2015zua,Geringer-Sameth:2015lua,2015ApJ...809L...4D,2016JCAP...02..039M,PhysRevD.93.043518}
and other promising sites such as the Galactic center
\cite{2011PhRvL.106p1301A,2011PhRvD..84l3005H,2015PhRvL.114h1301A,2015JCAP...03..038C,2016PDU....12....1D,2015PhRvD..91f3003C,2016ApJ...819...44A,Zhou:2014lva}
or clusters of galaxies
\cite{2010ApJ...710..634A,2010JCAP...05..025A,2012ApJ...750..123A,2012JCAP...07..017A,2012ApJ...757..123A,2015ApJ...812..159A,2016JCAP...02..026A,PhysRevD.93.103525}, both from the
ground with imaging Cherenkov telescopes and from space with the
\emph{Fermi} Large Area Telescope (LAT).
More recently, novel and competitive constraints have been obtained also
from the \emph{Fermi} measurements of the extragalactic gamma-ray background
\cite{2012PhRvD..85h3007A,2013MNRAS.429.1529F,2013PhRvD..87l3539A,2014PhRvD..90b3514A,2014NIMPA.742..149G,2015PhR...598....1F,
2015ApJS..221...29C,2015JCAP...06..029C,2015ApJ...802L...1F,Regis:2015zka,Ando:2016ang,Fornasa:2016ohl}.

In this paper, we focus on dwarf spheroidal galaxies (dSphs) that are
low-luminosity satellite galaxies which are known to be highly DM dominated
\cite{1998ARA&A..36..435M,2007PhRvD..75h3526S,2008ApJ...678..614S,2011JCAP...12..011S,Chiappo:2016xfs}.
Their high mass-to-light ratio, proximity, and very low expected
gamma-ray background from other astrophysical sources make them ideal
candidates to search for gamma rays from DM annihilation.
The main astrophysical uncertainty when dealing with indirect DM
searches in dSphs is their DM density profile, which is the most crucial
ingredient needed to estimate the rate of DM annihilation we expect from
a given object.
The common assumption often adopted in the literature is that dSphs
are characterised by a spherically symmetric, so-called
Navarro-Frenk-White (NFW) profile \cite{Navarro:1996gj}.
This cusped profile originally predicted by $N$-body simulations of cold
dark matter might not be the best choice for all cases, and other
profiles have been extensively discussed in the literature, including the
Einasto profile \cite{1965TrAlm...5...87E}. 

Additional complications come from going beyond simple spherical 
symmetric mass models. We know, in fact, that the observed stellar components of all MW 
dSphs have an axisymmetric shape on the sky-plane with typical axial ratios of 0.6--0.8
\cite{2012AJ....144....4M}. Additionally, recent high-resolution $N$-body simulations 
showed that DM subhalos tend to have axisymmetric shapes rather than triaxial
\cite{2014MNRAS.439.2863V}. These considerations prove the need to relax the 
assumption of spherical symmetry in the mass modeling of dSphs, which is also one 
of the major systematic uncertainties for the $J$-factor (i.e., the line-of-sight integral of 
DM density squared) estimations that most of previous studies have not considered.

In this paper we investigate the impact of observationally motivated axisymmetric mass models on
indirect DM searches with dSphs using gamma-ray observations by \emph{Fermi}. Uncertainties on the J-factor estimates were addressed in Ref.~\cite{Ullio:2016kvy}, where they explore the impact of the observationally unknown star orbital anisotropy.
Triaxial density profiles have been investigated in detail in
Ref.~\cite{Bonnivard:2014kza}, where they determine the bias on the
$J$-factor that arises when using a spherical Jeans analysis for halos
that are likely to be triaxial in shape.
In our work, we go beyond the $J$-factor estimates and 
study the impact on the upper limits obtained for the DM cross section
when adopting the axisymmetric models of Ref.~\cite{Hayashi:2015yfa} 
with respect to those obtained using the commonly adopted NFW profile.
We analyse about seven years of \texttt{PASS8} \emph{Fermi}-LAT data for seven
classical dSphs, namely Draco, Leo I and II, Sextans, Carina, Sculptor,
and Fornax.
These dSphs are selected as the overlapping part of the samples
considered by Ref.~\cite{Hayashi:2015yfa} and
Ref.~\cite{Ackermann:2015zua}.
We fit each dSph both with NFW and axisymmetric profiles, and compare
their cross section upper limits.
We underline, in particular, that Sextans, Carina, Sculptor and Fornax
are characterised by cored axisymmetric profiles rather than cusped, and
their results can differ significantly from those of the NFW profiles.

This paper is organised as follows.
In Sec.~\ref{sec:theomodel}, we discuss the expected flux from DM
annihilation from dSphs in the case of a NFW profile.
The axisymmetric mass model is introduced in Sec.~\ref{sec:axisym},
where we also discuss a qualitative comparison with the NFW profile.
In Sec.~\ref{sec:datasel}, we discuss the \emph{Fermi}-LAT data analysis for
the seven selected dSphs and present our results in
Sec.~\ref{sec:results}.
We discuss our conclusions in Section~\ref{sec:concl}.

\section{Gamma rays from dark matter annihilation}\label{sec:theomodel}

The gamma-ray intensity (i.e., the number of photons received per unit
area, time, energy, and solid angle) from a direction $\psi$ relative to
the center of the halo, expected from DM annihilation can be written as
\begin{equation}
\phi_{\text{WIMP}}(E,\psi) = J(\psi)  \Phi^{\text{PP}}(E) ,
\end{equation}
where $J(\psi)$ is the astrophysical factor, also called $J$-factor,
which describes the DM density distribution in the region of interest,
and $\Phi^{\text{PP}}(E)$ is the particle physics factor, which encloses
the properties of the DM particle.

The particle physics factor can be written as
\begin{equation}\label{eq:partfac}
\Phi^{\text{PP}}(E) = \frac{1}{2}\frac{\langle \sigma v \rangle}{4 \pi \, m^2_{\text{WIMP}}} \sum_f \frac{dN_f}{dE}B_f,
\end{equation}
where $m_{\text{WIMP}}$ is the WIMP mass, $\langle \sigma v \rangle$
is the the annihilation cross section multiplied by the relative
velocity of the annihilating particles averaged over their velocity
distribution, and $dN_f/dE$ is the photon spectrum of the final state
$f$ with its branching ratio $B_f$. 

The astrophysical $J$-factor is
\begin{equation}\label{eq:Jfac}
J(\psi) = \int_{\text{l.o.s.}} \rho^2(l,\psi) \, dl,
\end{equation}
%
where $l$ is the line-of-sight parameter, and $\rho(l,\psi)$ is the DM
density profile.
As mentioned in Sec.~\ref{sec:intro}, in our analysis of the \emph{Fermi}-LAT
data we compare the observationally-motivated axisymmetric DM density
profile with the widely used spherically-symmetric NFW profile.
The current section concerns the latter.

The NFW profile is given by \cite{Navarro:1996gj}
\begin{equation}\label{eq:NFW} \rho(r) = \left\{
  \begin{array}{lr}
    \frac{\rho_s r_s^3}{r(r_s+r)^2} &\text{for} \;\; r < r_t\\
    0 &\text{for} \;\; r \geq r_t 
  \end{array} ,
\right.
\end{equation}
where $\rho_s$ is the characteristic density, $r_s$ is the scale radius,
and $r_t$ is the tidal radius beyond which all the DM particles are
stripped away due to a strong tidal force from the host halo.
We calculate the values for $\rho_s$ and $r_s$ from the parameters
$v_{\text{max}}$ and $r_{\text{max}}$ provided by
\cite{2015MNRAS.451.2524M} using the following relations:
\begin{align}
r_s &= \frac{r_{\text{max}}}{2.163} \; , \\	
\rho_s &=  \frac{4.625}{4\pi G}\left( \frac{v_{\text{max}}}{r_s}\right)^2 ,
\end{align}
where $G$ is the gravitational constant.
We then derive $r_t$ from the Jacobi limit \cite{1987Binney}, 
\begin{equation}\label{eq:jacobi}
r_t = D\left(\frac{M_{\text{dSph}}}{3M_{\text{MW}}}\right)^{\frac{1}{3}},
\end{equation} 
where $M_{\text{dSph}}$ is the mass of the dSph and $D$ is the distance
of the dSph from the MW center.
$M_{\text{MW}}$ is the MW mass enclosed within the distance $D$,
calculated assuming an NFW profile from Ref.~\cite{Colafrancesco:2006he}.
$M_{\text{dSph}}$ is calculated integrating the dSph NFW profile up to
$r_t$, and we eventually solve equation~(\ref{eq:jacobi}) to obtain
$r_t$. Note that the tidal radius calculated in this way is subject to various uncertainties 
connected to the mass estimate of the Milky Way and to several assumptions made for simplicity, 
such as a perfect circular orbit of the dSph around the stable MW potential. 
However, typically about $90$\% of the annihilation flux comes from within $r_s$ for an NFW profile (see, e.g.,
\cite{2011JCAP...12..011S}) and, therefore, variations on the tidal radius will only have little 
effects on the resulting J-factor. The main characteristics of each considered dSph are
reported in Table~\ref{table:specs}.

\begin{table*}[hbt]
\caption{Characteristics of the analysed dwarf spheroidal galaxies. The top ones are cusped while bottom ones are cored in the axisymmetric mass modeling. The distances are taken from Ref.~\cite{Ackermann:2015zua}.}
\begin{tabular}{l*{8}{c}}
\hline 
\hline
Name		& Distance & \, $\rho_s$ & \,  $r_s$ & \, $r_t$ & \, NFW $J$-factor & & \, axisymmetric $J$-factor & \\
& & & & & total &   $<0.5^{\circ}$ & total & $<0.5^{\circ}$ \\

& [kpc]& [$M_{\odot}\text{kpc}^{-3}$] & [kpc] & [kpc] & [GeV$^2$
		     cm$^{-5}$ sr] & [GeV$^2$ cm$^{-5}$ sr] & [GeV$^2$ cm$^{-5}$ sr] &[GeV$^2$ cm$^{-5}$ sr] \\

\hline
Draco & 76 &$2.30\times 10^8$&0.3507&0.96& $8.33\times10^{18}$ & $8.24\times10^{18}$ & $9.43\times10^{18}$ & $8.71\times10^{18}$ \\
Leo I & 254 &$1.59\times 10^8$&0.4027&6.26& $5.06\times10^{17}$& $5.05\times10^{17}$& $3.95\times10^{17}$ &$3.94\times10^{17}$ \\
Leo II & 233 &$1.83\times 10^8$&0.3055&4.73& $3.32\times10^{17}$& $3.32\times10^{17}$ & $3.18\times10^{17}$ &$3.17\times10^{17}$ \\
\hline
Carina & 105 &$3.04\times10^8$&0.2065&3.39&$1.50\times10^{18}$& $1.49\times10^{18}$  & $2.61\times10^{18}$ & $2.36\times10^{18}$ \\
Fornax & 147 &$1.33\times 10^8$&0.4731&5.30& $1.83\times10^{18}$ & $1.81\times10^{18}$& $1.67\times10^{18}$ & $1.52\times10^{18}$\\
Sculptor & 86 &$1.67\times 10^8$ &0.3935&3.25& $4.91\times10^{18}$& $4.79\times10^{18}$ &$6.75\times10^{18}$ &$5.76\times10^{18}$\\
Sextans & 86 &$3.82 \times 10^8$&0.2018&1.55& $3.37\times10^{18}$& $3.37\times10^{18}$  & $2.03\times10^{18}$ &$1.24\times10^{18}$\\
\hline
\hline
\end{tabular}
\label{table:specs}
\end{table*}

Equation~(\ref{eq:Jfac}) yields the $J$-factor as a function of the
angle between the line of sight and the center of the dSph.
We project this onto a spatial map of $100 \times 100$ pixels of
$0.1^{\circ}$ centered on each dSph.
These will be the template input for the \emph{Fermi}-LAT data analysis of each
dSph that is described in Sec.~\ref{sec:datasel}.
We show the obtained NFW template maps in Figs.~\ref{fig:HalosCusped}
and \ref{fig:HalosCored}, where the total flux is normalized to unity.

Our reference works for the gamma-ray limits on dSph are those of
Refs.~\cite{Ackermann:2013yva,Ackermann:2015zua}.
However, while Refs.~\cite{Ackermann:2013yva,Ackermann:2015zua} 
limit their analysis within $0.5^\circ$ of each dSph, we do not limit the 
emission region in our analysis. Our choice is motivated by the fact that we want to 
compare the upper limits on the DM cross section obtained when adopting
the NFW profiles against those obtained when adopting axisymmetric ones.
As we will discuss in detail in the next section, the axisymmetric profiles are typically 
more extended compared to the NFW profiles. In Table~\ref{table:specs}, we
show the total $J$-factor together with the one calculated within a radius of 
$0.5^{\circ}$ both for the NFW and the axisymmetric profiles. While for the
NFW profiles the differences are insignificant, with maxima of about 1 and 2.5\% for 
Draco and Sculptor, respectively, the differences in the case of the axisymmetric mass
models are much more severe in most cases, except for Leo I and II. In particular, in the 
case of Sextans, about $40 \%$ of the total axisymmetric J-factor would be ignored by 
considering only a region within $0.5^\circ$. Therefore, in order to have a consistent comparison 
between the NFW and the axisymmetric profiles, we do not limit the emission region of our 
dSphs in the data analysis and use $r_t$ as outermost radius in the generation of 
the template input maps for the NFW case.

Finally, note that our NFW $J$-factors do not necessary have to coincide with
those of Refs.~\cite{Ackermann:2013yva,Ackermann:2015zua} as they use
the method of Ref.~\cite{2015MNRAS.451.2524M} applied to stellar
kinematics data to obtain their $J$-factors, while we use directly the
$v_{\text{max}}$ and $r_{\text{max}}$ provided by
Ref.~\cite{2015MNRAS.451.2524M}.
Nevertheless, our total $J$-factors integrated up to $r_t$ are always
within the quoted errors of the $J$-factors from
Refs.~\cite{Ackermann:2013yva,Ackermann:2015zua}, with the notable
exception of Leo~II where ours is almost a factor of 2 smaller.
With this exception in mind, we expect the limits that we calculate 
for NFW profiles to be comparable to those of 
Refs.~\cite{Ackermann:2013yva,Ackermann:2015zua}, except for the fact that here we consider events from 
a larger energy range.

\section{Axisymmetric Mass Models}
\label{sec:axisym}

Our aim is to compare the constraints obtained using an NFW density
profile to those obtained by using the observationally-motivated
axisymmetric density profile.
For the axisymmetric model, we use the non-spherical DM halo structure estimated by Ref.~\cite{Hayashi:2015yfa} to compute the $J$-factor
maps.
In this section, we briefly introduce the mass models based on the
axisymmetric Jeans equations, the method of exploring the best-fit DM
halo parameters, and the fitting results (for more details, we refer the
reader to the original papers~\cite{Hayashi:2015yfa,Hayashi:2016}).

Assuming that the stellar tracers in the dSphs are in dynamical
equilibrium with a gravitational smooth potential dominated by DM, the
distribution function obeys the steady-state collisionless Boltzmann
equation~\cite{2008gady.book.....B}.
Given that both the stellar and DM components are axisymmetric, the
axisymmetric Jeans equations can be derived from this equation by
computing its velocity moments.
When the distribution functions are of the form $f(E,L_z)$, where $E$
and $L_{z}$ are the energy and the angular momentum along the symmetry axis $z$ respectively, the mixed moments vanish and the velocity dispersion of
stars in cylindrical coordinates, $\overline{v^2_R}$ and
$\overline{v^2_z}$, are identical; i.e., the velocity anisotropy
parameter $\beta_z=1-\overline{v^2_z}/\overline{v^2_R}$ is exactly
zero.
However, since in general these velocity second moments are not
identical, Ref.~\cite{Hayashi:2015yfa} adopted Cappellari's formalism
that relaxed $\overline{v^2_R}=\overline{v^2_z}$ and assumed
$\beta_z={\rm constant}$~\cite{2008MNRAS.390...71C}.
In addition, they assumed that the dSph stars did not rotate, and therefore
the velocity second moment was equivalent to the velocity dispersion.

Under these assumptions, the axisymmetric Jeans equations are written as
\begin{eqnarray}
\overline{v^2_z} &=&  \frac{1}{\nu(R,z)}\int^{\infty}_z \nu\frac{\partial \Phi}{\partial z}dz,
\label{AGEb03}\\
\overline{v^2_{\phi}} &=& \frac{1}{1-\beta_z} \Biggl[ \overline{v^2_z} + \frac{R}{\nu}\frac{\partial(\nu\overline{v^2_z})}{\partial R} \Biggr] + 
R \frac{\partial \Phi}{\partial R},
\label{AGEb04}
\end{eqnarray}
where $\nu$ is the three-dimensional stellar density profile and $\Phi$ is the
gravitational potential.
In order to compare them with the observed velocity second moments, the
above equations should be integrated along the line of sight.
Following the method given in Ref.~\cite{2006MNRAS.371.1269T}, we
computed the projected velocity second moments from $\overline{v^2_R}$,
$\overline{v^2_{\phi}}$, and $\overline{v^2_z}$, taking into account the inclination
of each dSph with respect to the observer.
For the stellar and DM halo density models, which are related to $\nu$
and $\Phi$, we adopted an axisymmetric Plummer profile
\cite{1911MNRAS..71..460P} (see Eq.~3 in \cite{Hayashi:2015yfa}) and an
axisymmetric double power-law form (see Eq.~4 in
\cite{Hayashi:2015yfa}), respectively.

Comparing the line-of-sight velocity moment profiles from theory and
observations, Ref.~\cite{Hayashi:2015yfa} estimated the best-fit free
parameters by using a Markov Chain Monte Carlo fitting method.
There is a total of six free parameters in this model: the axial ratio,
characteristic density and scale radius of the DM halo, the inner slope
of the DM profile, the velocity anisotropy parameter and the inclination
angle of the dSph.
Applying their models to the available data of the seven MW dSphs
(Carina, Fornax, Sculptor, Sextans, Draco, Leo~I and Leo~II), two
important outcomes were found.
First, while Leo~I and Leo~II have almost spherical dark halos, the
other dSphs (Carina, Fornax, Sculptor, Sextans and Draco) 
are likely to have very flattened and oblate DM halos, 
with axial ratios of $\sim$0.4, even though there is a 
degeneracy between the axial ratio of the dark halo and the constant 
velocity anisotropy parameter. For example, the axisymmetric model 
for Sextans is preferred over a spherical symmetric one at around 
$2\sigma$ confidence level.
Second, not all the DM halos in the dSphs have a cusped central density profile.
Most of the dSphs indicate cored density profiles or shallow cusps.
Exceptions are Draco and Leo~I, which show a cusped profile with inner
density slopes of $-0.86\pm0.11$ and $-1.40^{+0.06}_{-0.08}$ respectively.
The best-fit parameters of each dSph are summarized in Table~2 of
Ref.~\cite{Hayashi:2015yfa}.
We use these parameters to compute the sky distribution of the
$J$-factors for Draco, Leo~I, Leo~II, Sextans, Carina, Sculptor and
Fornax.

\begin{figure}[t!]
 	\begin{subfigure}[b]{0.2\textwidth}
                \includegraphics[width=\textwidth]{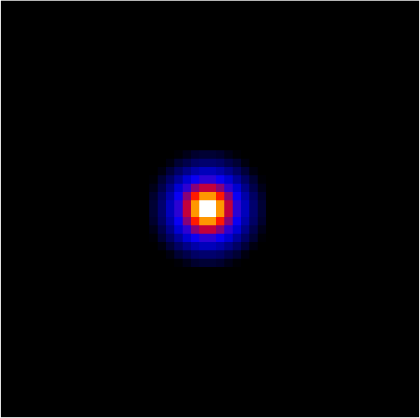} 
                \caption{\footnotesize Draco NFW}
                \label{fig:DracoS}
        \end{subfigure}
        \begin{subfigure}[b]{0.2\textwidth}
                \includegraphics[width=\textwidth]{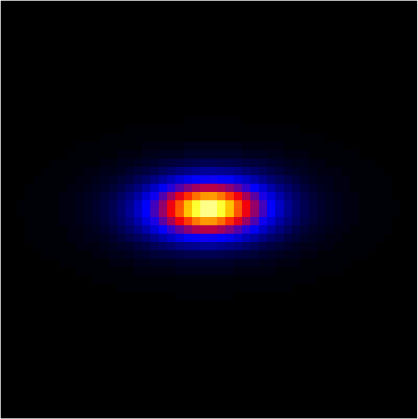}
                \caption{\footnotesize Draco axisymmetric}
                \label{fig:DracoAS}
        \end{subfigure}\\
        \begin{subfigure}[b]{0.2\textwidth}
                \includegraphics[width=\textwidth]{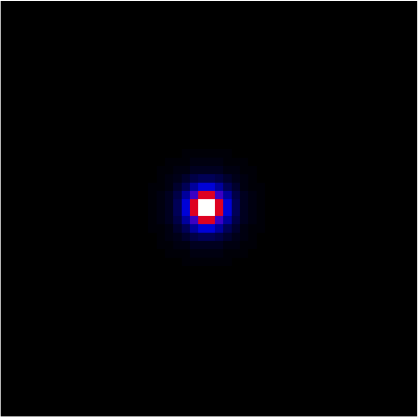}
                \caption{\footnotesize Leo I NFW}
                \label{fig:LeoIS}
        \end{subfigure}
        \begin{subfigure}[b]{0.2\textwidth}
                \includegraphics[width=\textwidth]{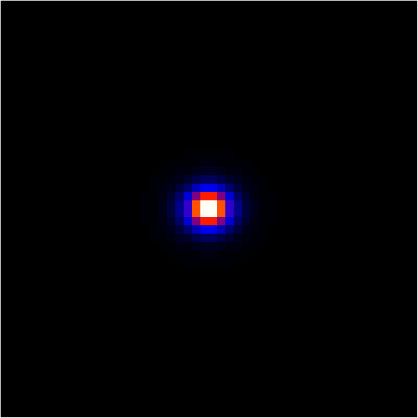}
                \caption{\footnotesize Leo I axisymmetric}
                \label{fig:LeoIAS}
        \end{subfigure}\\
        \begin{subfigure}[b]{0.2\textwidth}
                \includegraphics[width=\textwidth]{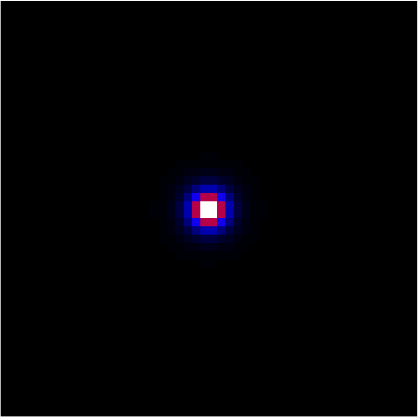}
                \caption{\footnotesize Leo II NFW}
                \label{fig:LeoIIS}
        \end{subfigure}
        \begin{subfigure}[b]{0.2\textwidth}
                \includegraphics[width=\textwidth]{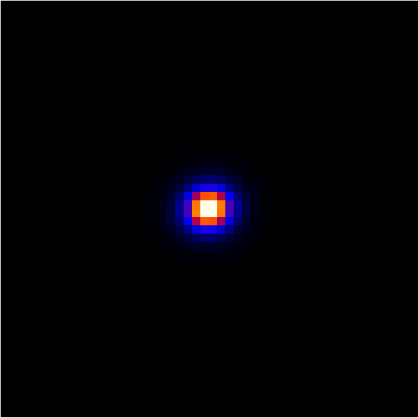}
                \caption{\footnotesize Leo II axisymmetric}
                \label{fig:LeoIIAS}
        \end{subfigure}
           \caption{DM density profiles projected onto the sky for the dSphs that have a cusped halo profile in log scale. From top to bottom, Draco, Leo~I and Leo~II, where the NFW profiles are shown on the left, and the axisymmetric profiles are shown on the right. The total flux of all images is normalised to unity, and the colour scale is the same in each pair of figures for every dSph. The maps are cropped to correspond to a $5^\circ\times5^\circ$ region in the sky.}\label{fig:HalosCusped}
\end{figure}

\begin{figure}[t!] 
 \centering   
        \begin{subfigure}[b]{0.2\textwidth}
                \centering
                \includegraphics[width=\textwidth]{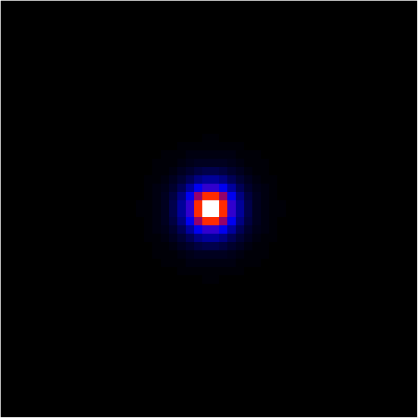}
                \caption{\footnotesize Sextans NFW}
                \label{fig:SextansS}
        \end{subfigure}
        \begin{subfigure}[b]{0.2\textwidth}
                \centering
                \includegraphics[width=\textwidth]{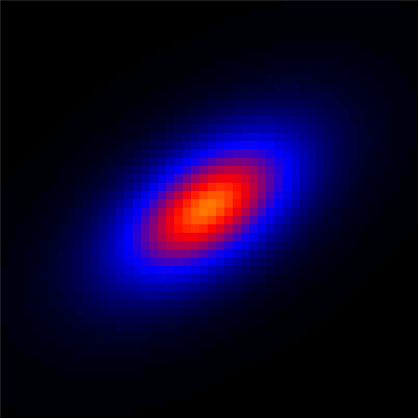}
                \caption{\footnotesize Sextans axisymmetric}
                \label{fig:SextansAS}
        \end{subfigure}\\
        \begin{subfigure}[b]{0.2\textwidth}
                \centering
                \includegraphics[width=\textwidth]{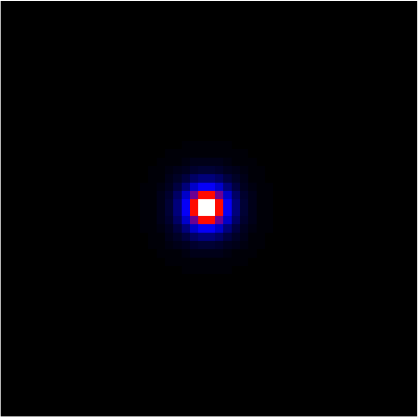}
                \caption{\footnotesize Carina NFW}
                \label{fig:CarinaS}
        \end{subfigure}
        \begin{subfigure}[b]{0.2\textwidth}
                \centering
                \includegraphics[width=\textwidth]{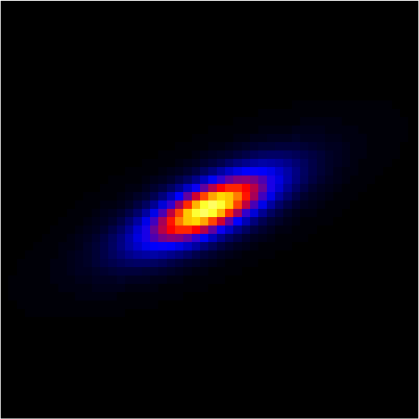}
                \caption{\footnotesize Carina axisymmetric}
                \label{fig:CarinaAS}
        \end{subfigure}\\
              \begin{subfigure}[b]{0.2\textwidth}
                \centering
                \includegraphics[width=\textwidth]{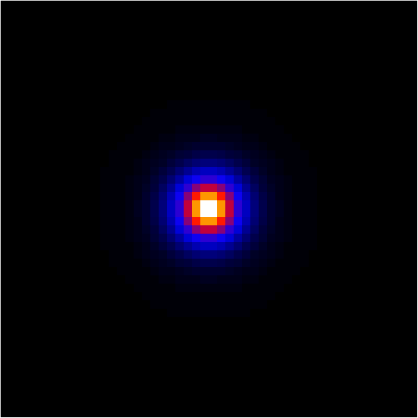}
                \caption{\footnotesize Sculptor NFW}
                \label{fig:SculptorS}
        \end{subfigure}
          \begin{subfigure}[b]{0.2\textwidth}
                \centering
                \includegraphics[width=\textwidth]{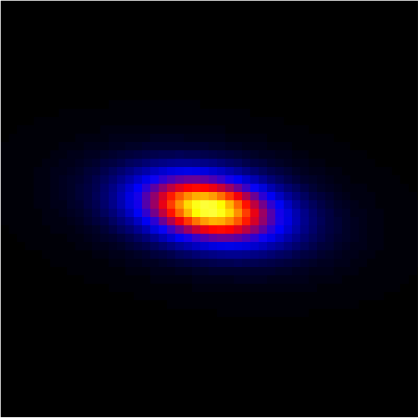}
                \caption{\footnotesize Sculptor axisymmetric}
                \label{fig:SculptorAS}
        \end{subfigure}\\
        \begin{subfigure}[b]{0.2\textwidth}
                \centering
                \includegraphics[width=\textwidth]{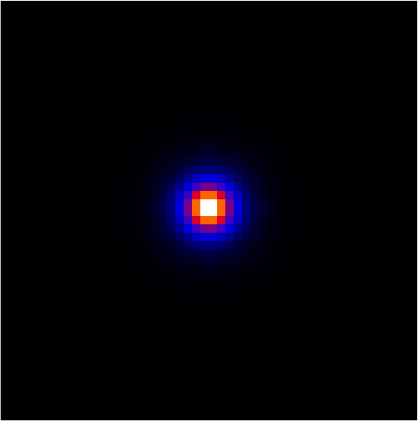}
                \caption{\footnotesize Fornax NFW}
                \label{fig:FornaxS}
        \end{subfigure}
        \begin{subfigure}[b]{0.2\textwidth}
                \centering
                \includegraphics[width=\textwidth]{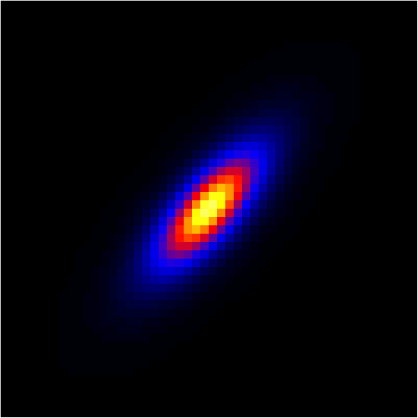}
                \caption{\footnotesize Fornax axisymmetric}
                \label{fig:FornaxAS}
        \end{subfigure}
        \caption{\small DM density profiles projected onto the sky for the dSphs that have a cored halo profile in log scale. From top to bottom, Sextans, Carina, Sculptor and Fornax, where the NFW profiles are shown on the left, and the axisymmetric profiles are shown on the right. The total flux of all images is normalised to unity, and the colour scale is the same in each pair of figures for every dSph. The maps are cropped to correspond to a $5^\circ\times5^\circ$ region in the sky.}\label{fig:HalosCored}
\end{figure}

Figures~\ref{fig:HalosCusped} and \ref{fig:HalosCored} show both the NFW
and axisymmetric density profiles projected onto the sky for the seven
adopted dSphs.These are the spatial templates that are used in the \emph{Fermi}-LAT data
analysis of Sec.~\ref{sec:datasel}.The total flux in these maps is normalised to unity, and the colour
scale of each pair NFW-axisymmetric is set to be the same, thus showing
the relative size and brightness of the two models for a given dSph.
As explained in the previous sections, when generating these template 
maps, the outermost radius is taken to be $r_t$ for the case of the NFW profiles. 
In the case of the axisymmetric profiles, for which $r_t$ values are not estimated 
within the framework of Ref.~\cite{Hayashi:2015yfa}, there is no formal limit to 
the radial extent of the profiles in the template maps. We stress, however, that 
as in both cases most of the annihilation flux comes from the inner parts, even 
though with the due differences (see Table~\ref{table:specs}), the choice of the 
outermost radial extent has no impact on our results.

Figure~\ref{fig:HalosCusped} shows the dSphs with a cusped density
profile.
For Leo~I and Leo~II, there is almost no visible difference between the
NFW and the axisymmetric profiles projected onto the sky.
For Draco, the shape of the axisymmetric model is oblate instead of
spherical and clearly differs from the classical NFW, but still shows a cuspy.
The differences between the two profiles are larger for the cored dSphs
as can be seen in Fig.~\ref{fig:HalosCored}.
In this case, the axisymmetric profiles are much more extended than the
NFW profiles, with the total integrated $J$-factor being of the same
order of magnitude, but distributed over a larger area (see also Table~\ref{table:specs}).
Note also that these axisymmetric profiles are all oblate and
characterised by different directions of the major axis following the
stellar kinematics data for a given dSph.
We will show that the case of the cored dSphs is the most affected by
the simplification of adopting the NFW profile when obtaining DM
constraints.

\section{Data Selection and Analysis}
\label{sec:datasel}

We analyse $86$~months (August 4th 2008 15:43:36 till October 15th 2015
02:34:52) of \emph{Fermi}-LAT \texttt{PASS8} data in the direction of the
selected dSphs using the \texttt{v10r0p5} version of the \emph{Fermi}
\emph{Science Tools}.
We follow Ref.~\cite{Ackermann:2015zua} for the selection of event class
and type (\texttt{evclass}=128, \texttt{evtype}=3) and for the data
cuts, which are standard, and use the corresponding instrumental
response functions.
We analyse a region of interest (ROI) of $10^{\circ}\times10^{\circ}$
around each dSph, with $0.1^{\circ}$ pixels, and perform a binned
likelihood analysis in 24 logarithmically-spaced energy bins from
$100$~MeV to $50$~GeV.

We perform the analysis including all the sources included in the third \emph{Fermi}
catalog (3FGL; \cite{Acero:2015hja}) within a region with a radius of
$25^{\circ}$ around the center of our ROI for each dSph.
For the diffuse background, we adopt the latest Galactic diffuse model
(\texttt{gll$\_$iem$\_$v06}) and the extragalactic isotropic diffuse
model (\texttt{iso$\_$P8R2$\_$SOURCE$\_$V6$\_$v06}) as provided by the
\emph{Fermi} collaboration.
We allowed the spectral parameters of the sources to vary within a
circle of radius $7.07^{\circ}$---the radius of our ROI---together with
the normalisation of the diffuse background components, while the
remaining sources are kept fixed to the 3FGL values.

The so-obtained model is complemented in each case with the spatial
models of Figs.~\ref{fig:HalosCusped} and \ref{fig:HalosCored} for the
dSphs' DM-induced emission.
For each dSph, we run two separate analyses with the
corresponding NFW and axisymmetric profiles.
The spectral part of our dSphs' models is constructed using
Eq.~(\ref{eq:partfac}) adopting the corresponding J-factor for the NFW
or axisymmetric model from Table \ref{table:specs}, and making a guess
for the value of $\langle \sigma v \rangle$---the parameter that we will
constrain.
As for the photon spectrum $dN_f/dE$, we adopt {\tt PYTHIA}~\cite{PYTHIA}
for the $b\bar{b}$ final state.
The normalisation of our dSphs' models is left free.
In each case, we repeat the analysis for 18 values of the DM masses from
10 to 5000~GeV.

We run the binned likelihood analysis following the above prescriptions
for each dSph, for both a NFW and an axisymmetric profile, and for each
DM mass.
When convergence is not achieved, we iterate by filtering out the
faintest sources in our model with test statistic (TS) values $\leq 1$,
and subsequently $\leq 2$, while making sure that the model is still a
good description for the data.
We eventually calculate $95$\% confidence-level integrated flux upper
limits between 100~MeV and 50~GeV for all cases and derive limits on the
DM annihilation cross section that we discuss in detail in the next
section.

\begin{figure}[t!]
 	\begin{subfigure}[b]{0.19\textwidth}
                \includegraphics[width=\textwidth]{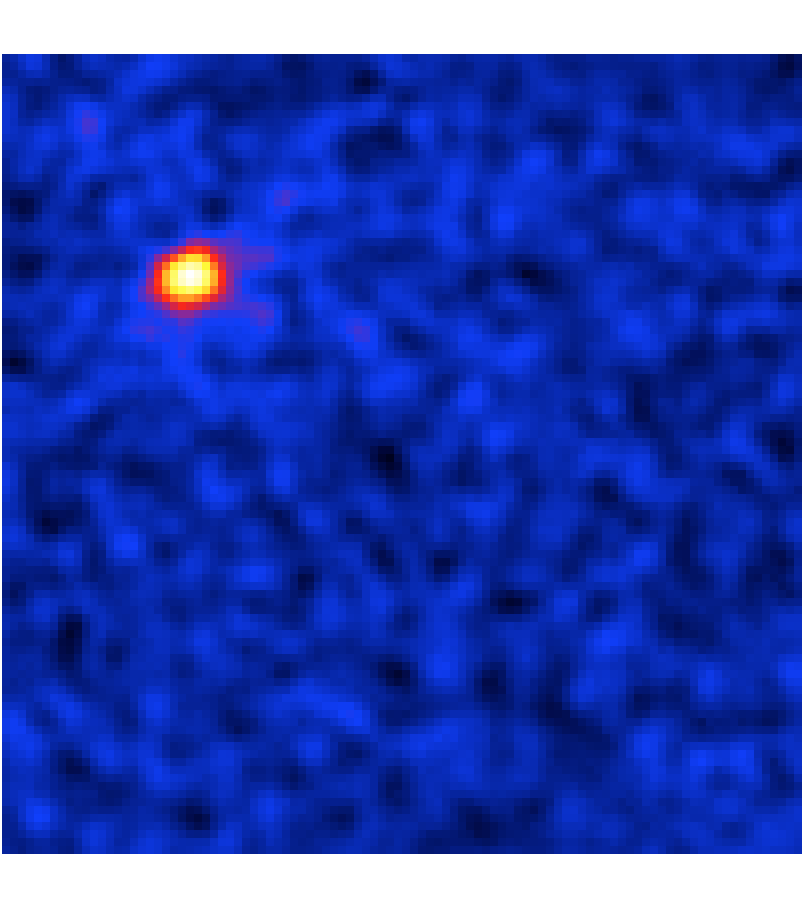} 
        \end{subfigure} 
        \begin{subfigure}[b]{0.19\textwidth}
                \includegraphics[width=\textwidth]{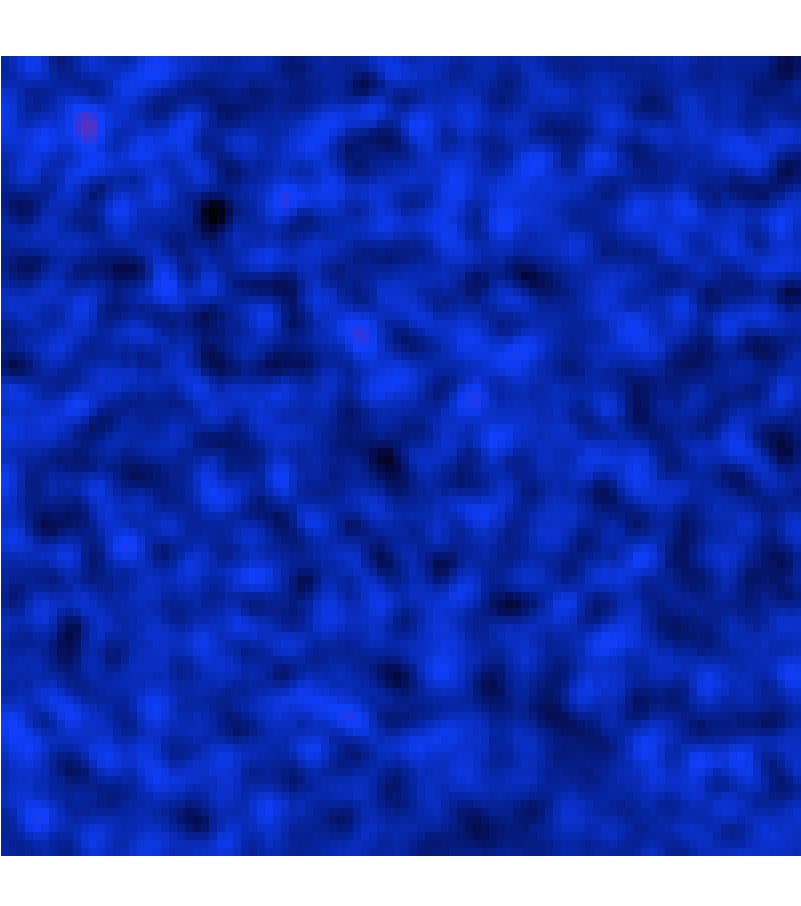} 
        \end{subfigure}
          \begin{subfigure}[b]{0.05\textwidth}
                \includegraphics[width=\textwidth]{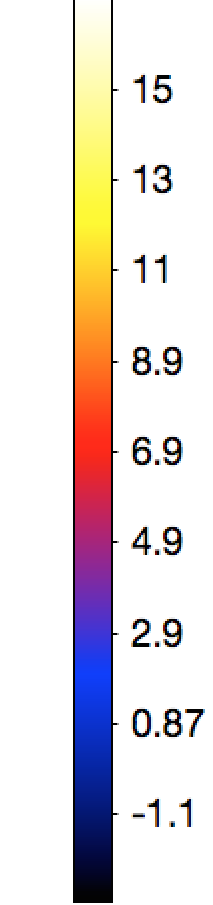} 
        \end{subfigure}
       \caption{Maps for Sextans, covering $10^\circ \times 10^\circ$ of the sky. On the left is the residual map before modeling the source, on the right the residual map after modeling the source. The residual maps represent subtractions of the model map from the counts map, therefore the color code refers to residual photon counts.}\label{fig:residuals}
\end{figure}

Before we move on to the results, we comment on the model used for the
analysis of Sextans.
The residual map for Sextans showed the presence of an unmodeled excess
at about $3.5^\circ$ from the center of the ROI, as shown in
Fig.~\ref{fig:residuals}, for which we did not find any correspondence
in the 3FGL catalog or in the literature.
The position of this excess is roughly $(155.93 , 0.65)$ in celestial
coordinates.
We fit this excess with a point source described by a simple power law
spectrum.
We found that this source had a TS value around 1460 and its spectrum
was well described by $dN/dE = 12.14 \times 10^{-9}\left(E/28.04~{\rm
MeV}\right)^{-2.39} \mathrm{cm^{-2}~s^{-1}~{MeV}^{-1}}$, with normalisation and spectral index having variations
below $1\%$ among the various analyses we ran for the NFW and
axisymmetric profiles and different DM masses.
In Fig.~\ref{fig:residuals} we show the residual map before and after
including this source for one of the analyses. We do not attempt any
further modeling or interpretation for this excess, considering our fit
just an effective model for it.
We are confident that this is a good description of the data for the
purposes of our work, also because the derived upper limits on the
annihilation cross section from Sextans differ very little if we do or
do not model this excess out from the data. Nevertheless, the results 
that we will discuss in the following section refer to the case where 
we model this source out.

Note that the \emph{Fermi} Collaboration published results using 
energies from 500~MeV to 500~GeV \cite{Ackermann:2013yva,Ackermann:2015zua}
while we use the 100~MeV to 50~GeV energy range. In particular, Ref.~\cite{Ackermann:2015zua} 
excluded events below 500~MeV to mitigate the impact of leakage from the bright limb of the Earth.
As the same analysis chain is applied to both profile types, the choice of the energy range do not 
impact the conclusions of our work, i.e., the comparison of the exclusion limits on the DM cross 
section between NFW and axisymmetric profiles. To confirm this, we perform the analysis of 
Sextans, which, as will be discussed in the next section, shows the largest difference between 
the two models, also in the energy range between 500~MeV to 50~GeV. The results are shown in
the top right panel of Figure~\ref{fig:resultscored}. As expected, the limits improve when excluding
lower energy events from our analysis, particularly for low DM masses. The limits are consistently
better for all tested DM masses in the case of a NFW profile, while in the case of the axisymmetric 
profile, the limits obtained in the 500~MeV$-$50~GeV range slightly worsen for DM masses above
about 100~GeV. At any rate, the relative comparison between the constraints obtained with NFW and
axisymmetric profiles is not affected by the choice of the energy range.

\section{Results}\label{sec:results}

We find no gamma-ray excess in any of the dSphs using both the NFW
profile and the axisymmetric models.
For most of the dSphs and DM masses, we find test statistic (TS) values
around zero, and no TS values were larger than $6.06$, which was the
case for Fornax using the axisymmetric profile (5.6 using the NFW
profile) and a DM mass of $m_{\rm WIMP} = 10 ~{\rm GeV}$.
Therefore we calculate flux upper limits that we then convert to limits
on the annihilation cross section.

\begin{figure*}[t!]
	\centering
 	\begin{subfigure}[b]{0.49\textwidth}
                \includegraphics[width=\textwidth]{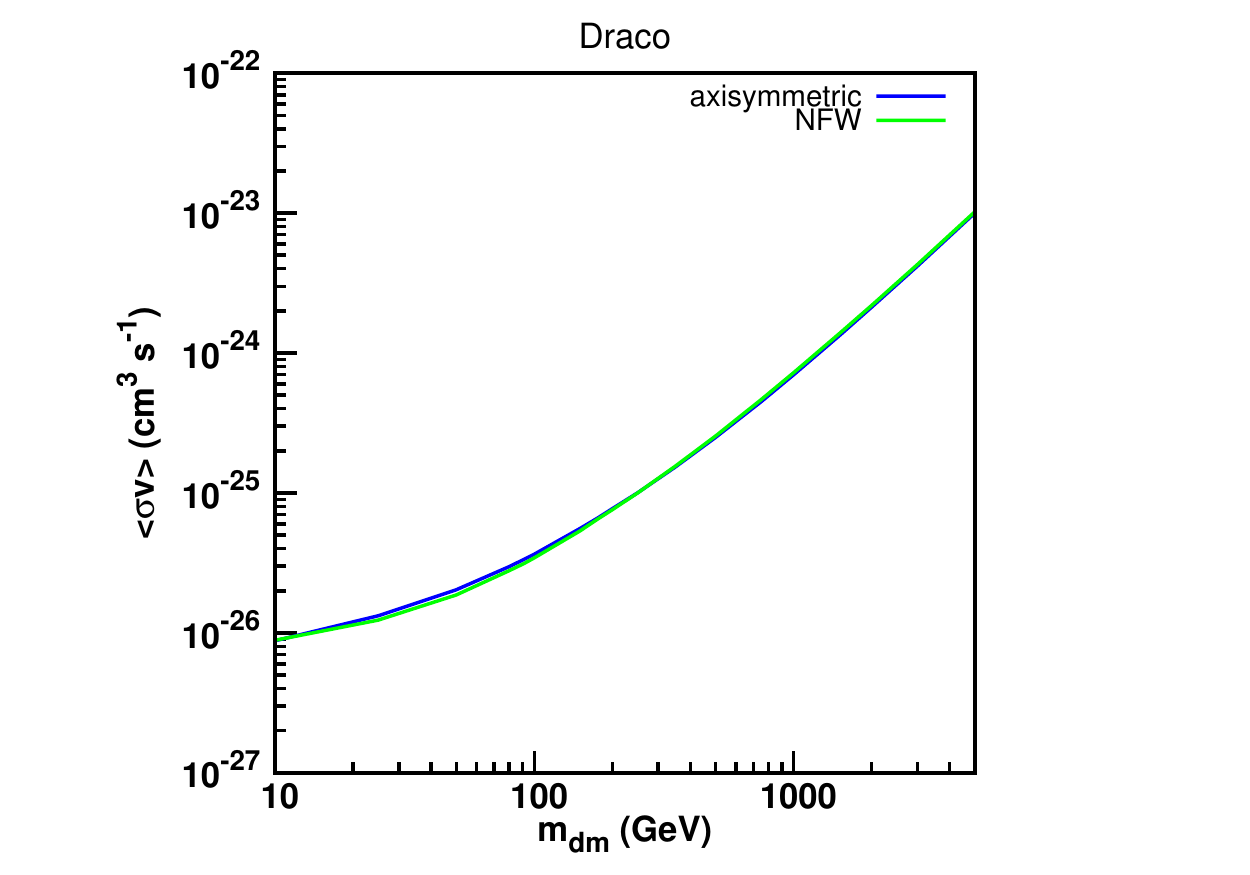} 
       \end{subfigure}
        \begin{subfigure}[b]{0.49\textwidth}
                \includegraphics[width=\textwidth]{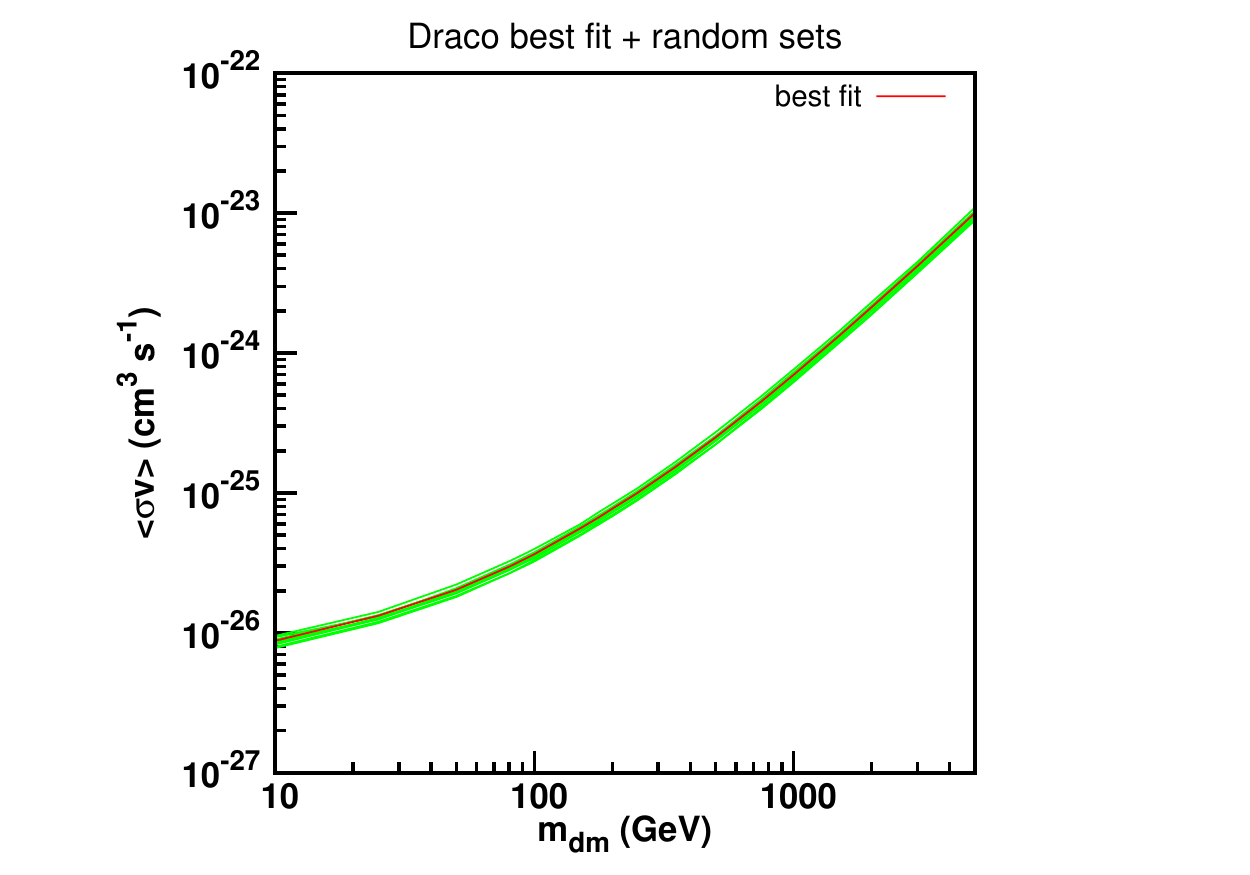}
       \end{subfigure}\\
        \begin{subfigure}[b]{0.49\textwidth}
                \includegraphics[width=\textwidth]{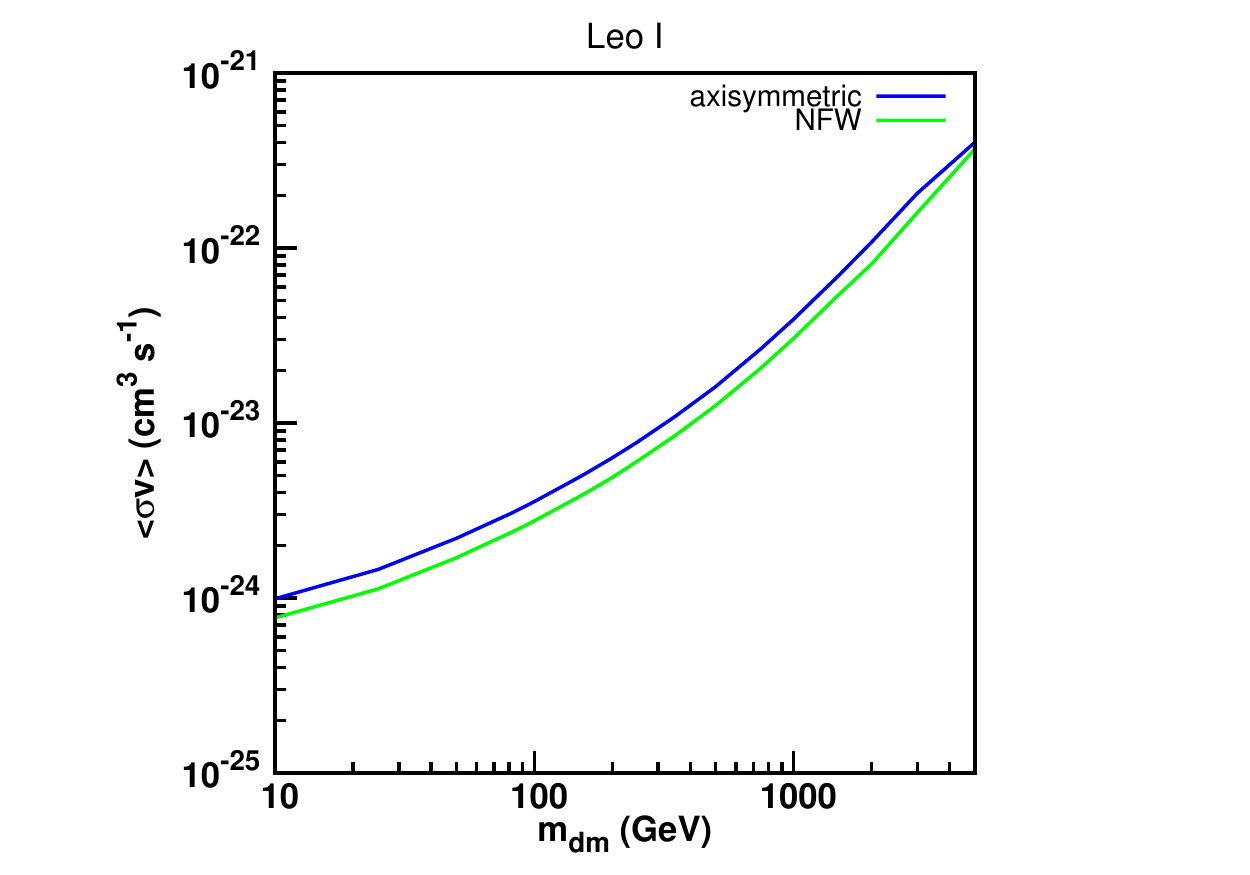}
       \end{subfigure}
        \begin{subfigure}[b]{0.49\textwidth}
                \includegraphics[width=\textwidth]{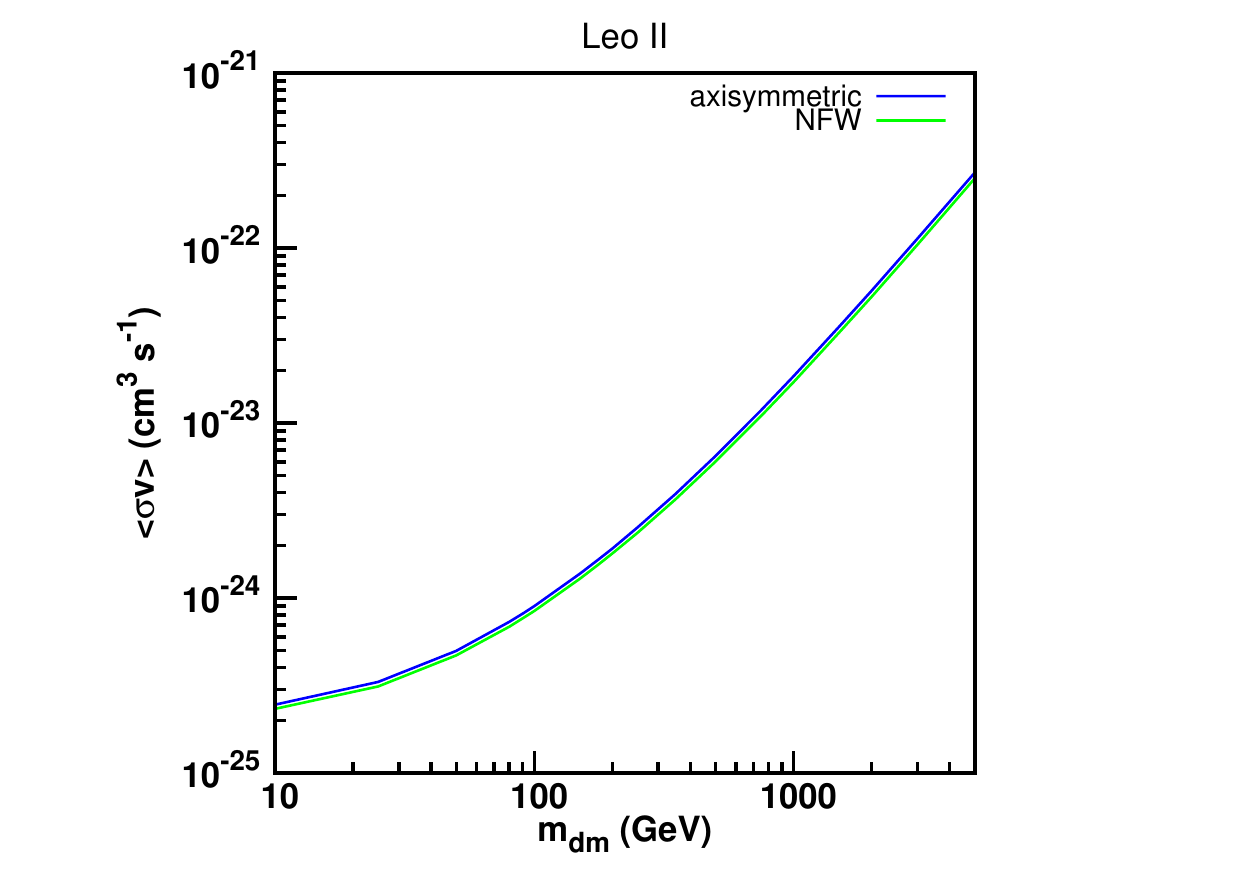}
       \end{subfigure}\\
        \caption{Dark matter annihilation cross section upper limits in
 the $b\bar{b}$-channel for the dSphs with a cusped profile. The
 upper right frame shows the cross section upper limits obtained
 through the analysis of 10 axisymmetric profiles for Draco,
 corresponding to 10 random sets of the profile parameters from the
 Monte Carlo sample of Ref.~\cite{Hayashi:2015yfa}, along with the
 best-fit case.  }\label{fig:resultscusped}
\end{figure*}

\begin{figure*}[t!]
	\centering
 	\begin{subfigure}[b]{0.49\textwidth}
                \includegraphics[width=\textwidth]{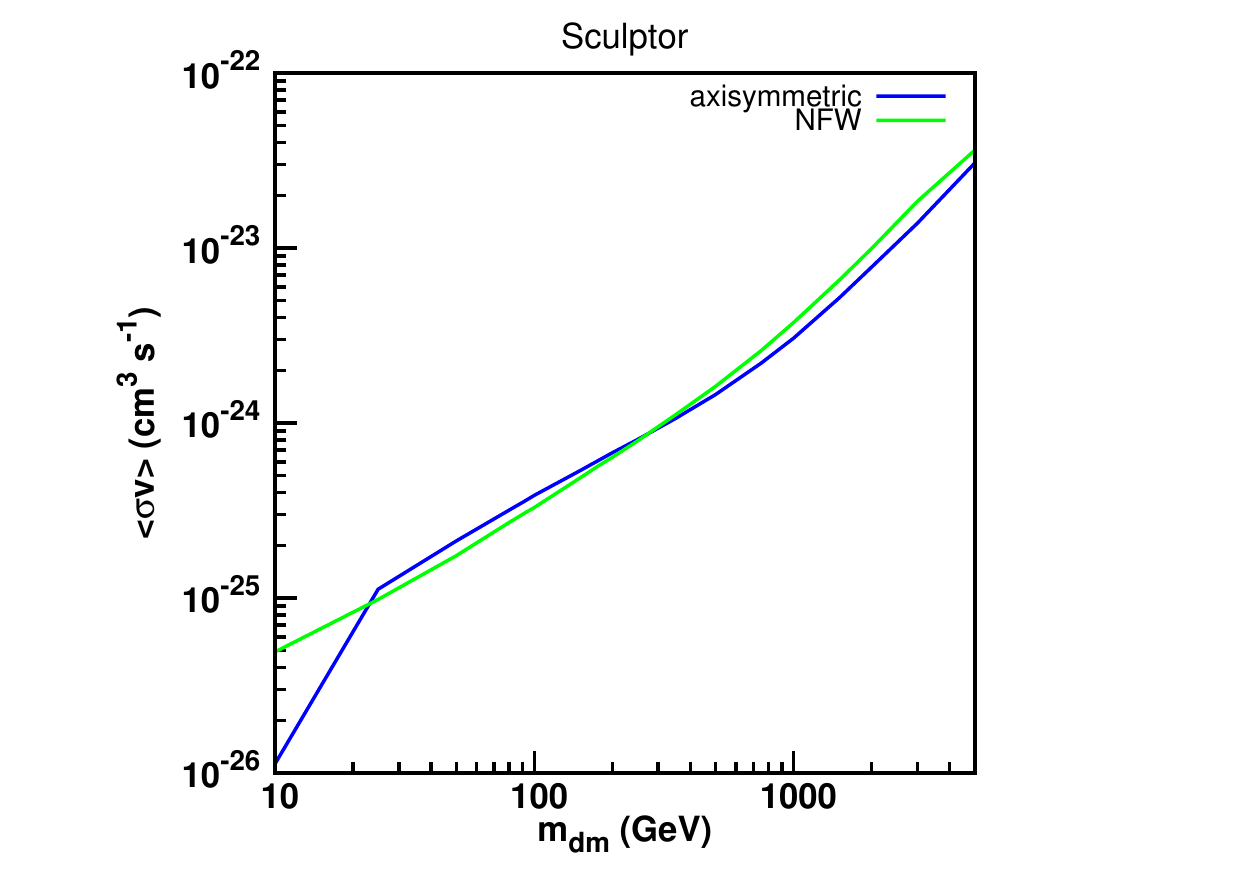} 
        \end{subfigure}
        \begin{subfigure}[b]{0.49\textwidth}
                \includegraphics[width=\textwidth]{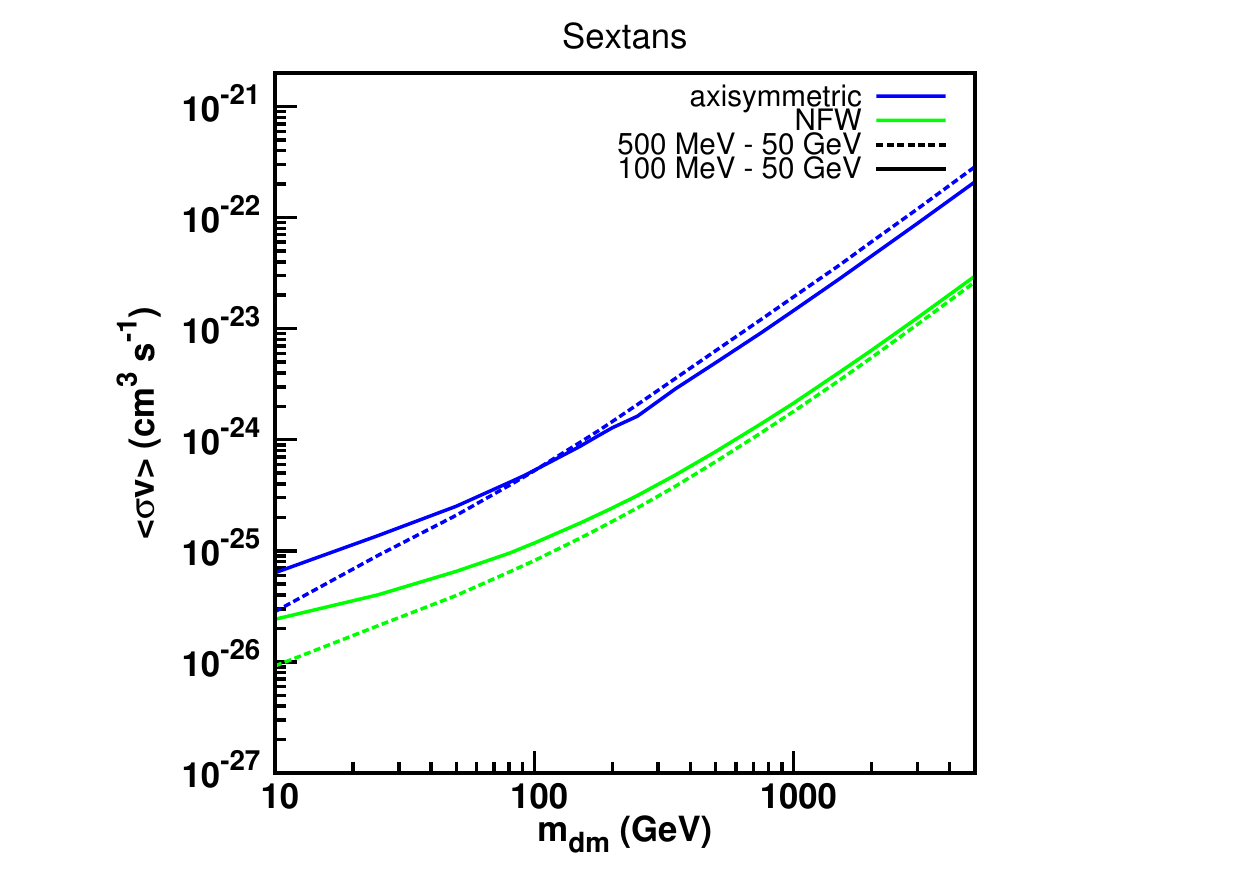}
        \end{subfigure}\\
        \begin{subfigure}[b]{0.49\textwidth}
                \includegraphics[width=\textwidth]{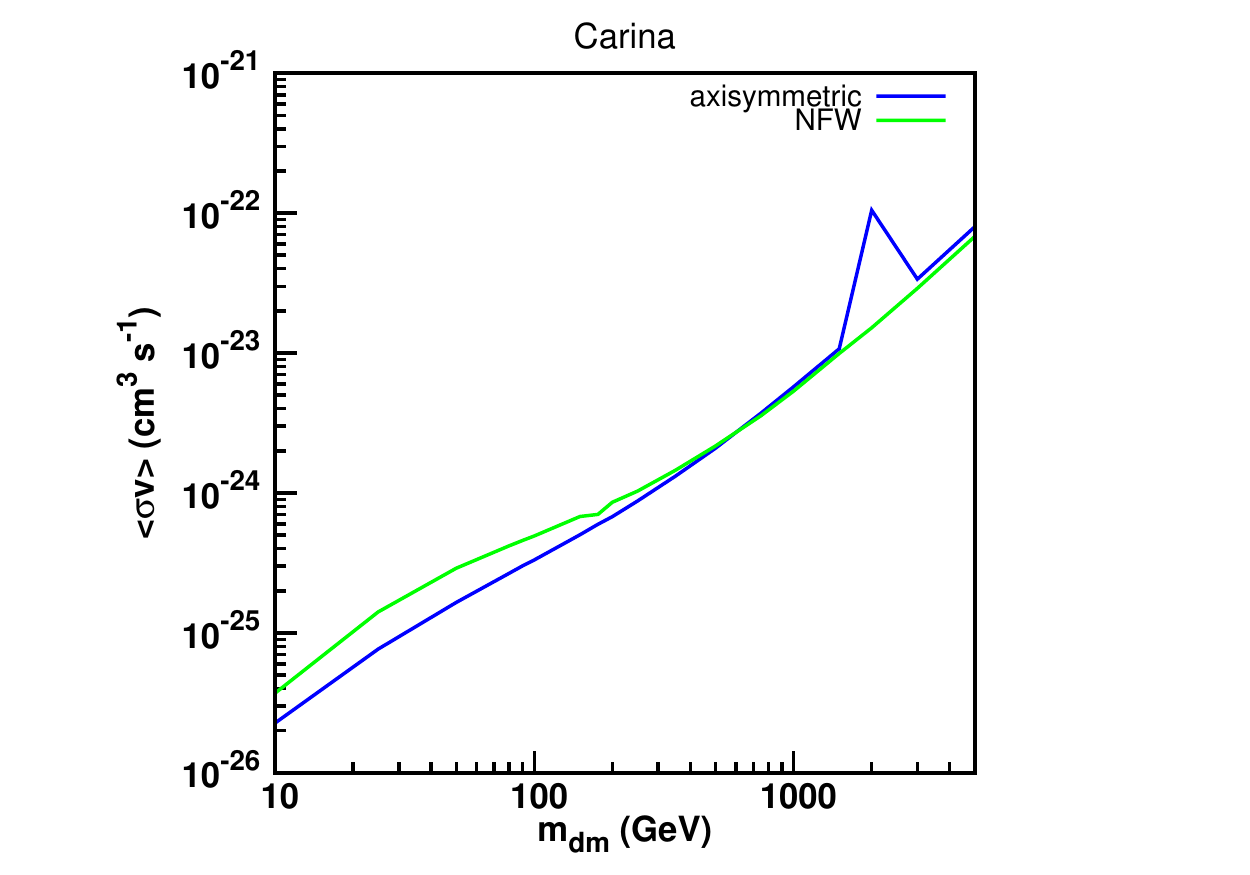}
        \end{subfigure}
        \begin{subfigure}[b]{0.49\textwidth}
                \includegraphics[width=\textwidth]{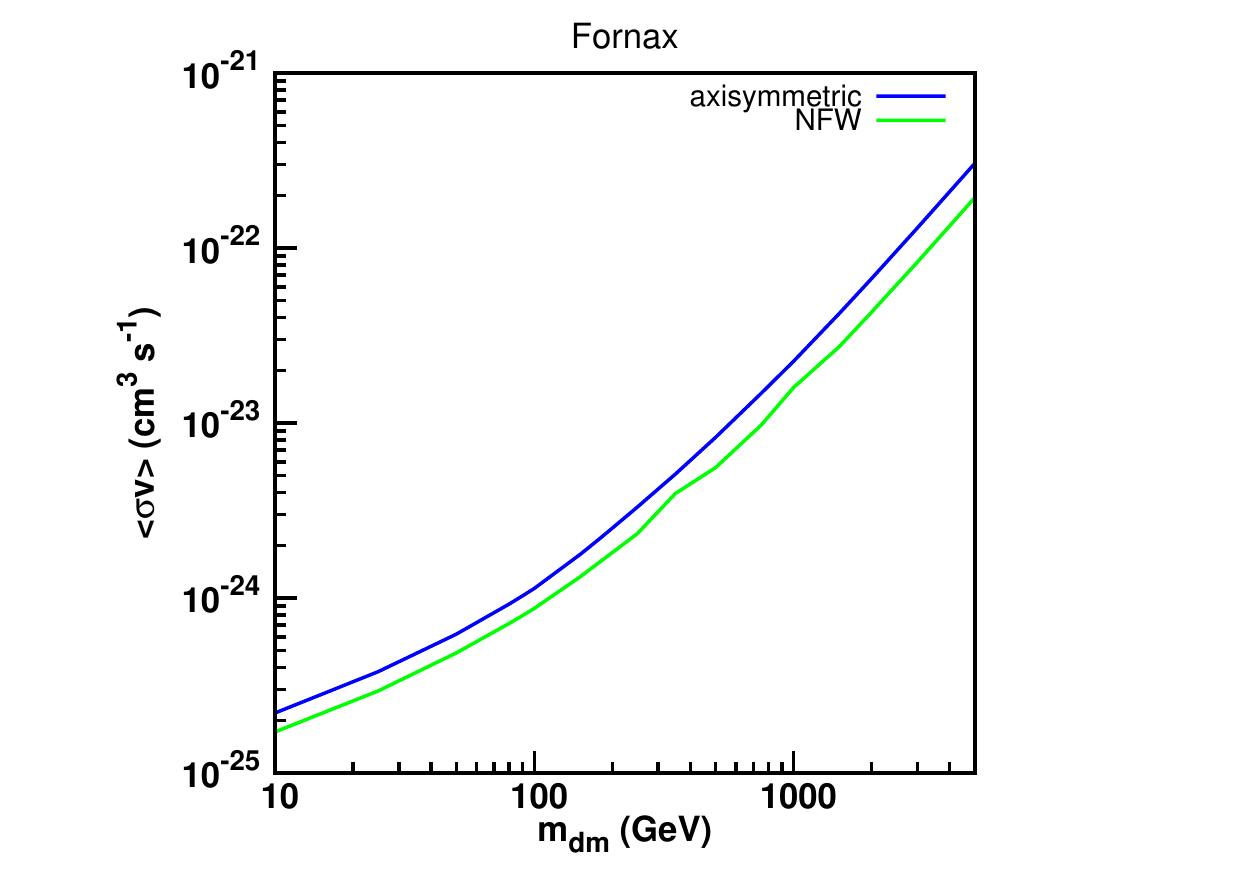}
        \end{subfigure}\\
        \caption{Dark matter annihilation cross section upper limits in the $b\bar{b}$-channel for the dSphs with a cored profile.}\label{fig:resultscored}
\end{figure*}

We find differences between the cross section upper limits achieved
through the two different models of the halo profile.
In Figs.~\ref{fig:resultscusped} and \ref{fig:resultscored}, we show the
cross section upper limits for the seven analysed dSphs.
Figure~\ref{fig:resultscusped} shows that the dSphs that are expected to
have a cusped profile show small differences in the upper limits for the
two analysed halo models.
Despite the difference in the shape of the two halos (spherical vs
oblate), we find that the NFW profile provides a good approximation of
the actual halo of these dwarfs.

The impact of the different profiles is more significant for the
four dSphs that have a cored profile as suggested by the
observationally motivated profile we adopted and shown in
Fig.~\ref{fig:resultscored}.
In particular, we find the largest difference of about a factor of
2.5--7, depending on the DM-particle mass, in the case of Sextans, where
we see that the axisymmetric model is most extended compared with the
corresponding NFW profile as shown in Fig.~\ref{fig:HalosCored}.
The upper right panel in Fig.~\ref{fig:resultscored} shows 
the resulting cross section upper limits of Sextans derived both in the energy 
range 100~MeV$-$50~GeV and 500~MeV$-$50~GeV. As anticipated in the 
previous section, the difference between the NFW and axisymmetric profiles 
is unaffected by this choice. It is, in fact, even slightly larger -- a factor of 3--11 depending 
on the DM-particle mass -- for the 500~MeV$-$50~GeV energy range.
Therefore, given that Sextans was one of the most important dSphs with the
spherically symmetric model, i.e., cross section upper limits reached
the canonical value $3\times 10^{-26}~\mathrm{cm^{3}~s^{-1}}$ for
low-mass WIMPs, we show that it is indeed relevant to use a more 
accurate model for its density profile.

The most stringent constraints on $\langle \sigma v \rangle$ are
obtained for Draco, whose $J$-factor is the largest among the seven
dSphs analyzed here.
In this case, the canonical annihilation cross section $\langle \sigma v
\rangle = 3 \times 10^{-26}~\mathrm{cm^{3} ~ s^{-1}}$ can be tested for
WIMPs lighter than $\sim$80~GeV, and since the DM density is
described by the cusped profile, there is only little difference between
the spherical and axisymmetric models.
Although the results of the combined likelihood analysis (e.g.,
Ref.~\cite{Ackermann:2015zua}) will be dominated by the most promising
dSphs such as Draco, others, such as Sextans discussed above, will also
give a substantial contribution.
Therefore, the inclusion of observationally-motivated axisymmetric 
profiles would make the joint likelihood analysis of the dSphs slightly weaker 
compared to the previous analysis in the literature.

To test the impact of measurement uncertainties of stellar kinematics
data on these gamma-ray constraints, we randomly choose ten sets of the
parameters from the Monte Carlo sample of Ref.~\cite{Hayashi:2015yfa} for 
the Draco axisymmetric profile, and obtain the cross section upper limits for each, 
whose results are shown in the upper right panel of Fig.~\ref{fig:resultscusped} along with the
best-fit case. This shows that the current stellar kinematics data are well determined,
giving only uncertainties on the cross section upper limits of about
10\%, which makes dSphs a robust, and hence, attractive object to test
DM annihilation.\footnote{We generated 100 random 
sets from the Monte Carlo sample of Ref.~\cite{Hayashi:2015yfa} for the
Draco axisymmetric profile. We then randomly select only 10 of these on which
to run our \emph{Fermi} analysis for each DM mass, as this can be a quite
lengthy process. We note, however, that the difference in the total $J$-factors
of the original 100 sets is within few percent at most. Therefore, we believe that
our choice of running only 10 sets provided robust results.}
This also shows that our comparison between NFW and
axisymmetric profiles is not significantly affected by the uncertainties on the latter 
and that our conclusions are robust.

We note that a kink around $\sim$2~TeV for the axisymmetric model of
Carina as well as a drop toward $\sim$10~GeV of Sculptor is likely
caused by some complicated interplay between the adopted profile, energy
spectrum, and photon count distribution that we interpret as a statistical 
fluctuation, also considering that the models for these particular cases
of $m_\mathrm{dm}$ show no substantial difference, i.e., in TS
significance, with respect to the others.

Finally, we note that although evaluating the integrated $J$-factor will
capture the overall importance of each dSph, it is not until one performs
the likelihood analysis that we know how the cross section upper limits
behave as a function of the WIMP mass.
In fact, the difference in the cross section upper limits comes from an
interplay of the normalisation and shape of the $J$-factor.
For example, the difference between the $J$-factors is larger for Leo I
than for Fornax, with a value of 0.78 against 0.91 for the ratio
$J_{\text{axisymmetric}}/J_\mathrm{NFW}$.
The difference between the upper limits however is larger for Fornax,
where the upper limit for the axisymmetric case is up to 1.57 times
larger than the NFW case, while up to 1.33 times larger in the case of
Leo I.
So the difference between the shapes of the halo models has a larger
contribution to the difference in the cross section upper limits than
the difference between the total $J$-factors. 
While Ref.~\cite{Bonnivard:2014kza} studied $J$-factors for a
comprehensive list of dSphs, our focus is on the classical seven dSphs
that have the best measurements of stellar kinematics, and we performed
the likelihood analysis for all of them.
Therefore, these two approaches are complementary to each other.

Before moving to the conclusions, we want to underline that the cross section upper limits shown here, differently from Refs. \cite{Ackermann:2013yva,Ackermann:2015zua}, are obtained without taking in consideration any uncertainty, i.e., without marginalising over the uncertainty on, e.g., the J-factor determination. However, this has no impact on the relative comparison that we set out to make between the NFW and axisymmetric profiles.

\section{Conclusion}\label{sec:concl}

Dwarf spheroidal galaxies are important and well established targets for
indirect DM searches.
The most common choice for the DM density profile in the analysis of
these dSphs is an NFW profile.
Recent observational data of stellar kinematics, however, imply that DM
halos around these galaxies are better described by an axisymmetric
profile, with an axis ratio of 0.6--0.8, either cored or cusped.
For this reason, we investigated the impact of adopting
observationally-motivated axisymmetric models instead of the 
commonly adopted NFW profile on the limits obtained for the DM annihilation 
cross section for seven classical dSphs with \emph{Fermi} gamma-ray data.

Draco is the most promising dwarf galaxy among the seven analysed.
Although its DM distribution is well described by a cusped oblate
profile in the axisymmetric modeling, the total amount of gamma rays yielding 
from the overall region will be similar to that of an NFW profile (i.e., similar $J$-factors).
As a result, we obtained very similar upper limits on the annihilation
cross section for Draco using an NFW and axisymmetric model.
The same is true for Leo~II, while Leo~I shows some mild differences,
even if both feature an inner cusp.
By testing ten axisymmetric profiles randomly chosen from 
a Monte Carlo sample of the analyses of stellar kinematics data of Draco, 
we find that the current uncertainty on the density profile of Draco will give
a systematic uncertainty on the cross section upper limits of about
10\%. This proves that our conclusions are robust.

The analyses of the dSphs best described by a cored profile (Sextans, Sculptor, 
Carina and Fornax) result in a more substantial difference between the two adopted 
profiles. In particular, for Sextans, the best-fit model of its stellar kinematics
data yields a much more extended $J$-factor map.
We found that the cross section upper limits were weaker
by a factor of a few to several compared with those obtained with an NFW
profile. This demonstrates the importance of properly assessing DM
density profiles from observational data, and also that upper limits in
the literature obtained assuming a cusped spherical model (such as an NFW)
might be overestimated.

\acknowledgments
This work was supported by the Foundation for Fundamental Research on Matter
(FOM) through the FOM Program (N.K. and S.A.), the Dutch Organization for
Scientific Research (NWO) through Veni (F.Z.) and Vidi (S.A.) grants,
and partly by the Japan Society for the Promotion of Science~(JSPS)
KAKENHI Grant Number 16H01090 (K.H.).
We also thank Stephan Zimmer for the useful discussions, and the anonymous referee for the helpful comments.

\bibliographystyle{h-physrev}
\bibliography{mybib}

\end{document}